\documentclass[checkin,showpacs,aps,pra]{revtex4-1}

\newcommand{\bn}{\begin{enumerate}}
\newcommand{\en}{\end{enumerate}}
\newcommand{\ba}{\begin{eqnarray}}
\newcommand{\ea}{\end{eqnarray}}

\usepackage{graphicx,color}

\newcommand{\be}{\begin{equation}}
\newcommand{\ee}{\end{equation}}

\newcommand{\et}{{\it et al. }}
\newcommand{\ete}{{\it et al.}}

\def\prl{{ Phys. Rev. Lett. }}

\def\prb{{ Phys. Rev. B }}

\begin{document}

\newcommand{\clr}{\color{black}}

%\title{Laser-induced spin-orbit torque controls spin switching in
%  ferromagnetic nanostructures: Rise of orbital angular momentum }

%\title{ Femtosecond spin-orbit torque heralds all-optical spin
%  switching in ferromagnets: Rise of orbital angular momentum }

%\title{ Switching 40,000 ferromagnetic spins by an ultrafast laser
%  pulse:\\ Emergence of optical spin-orbit torque }

\title{Is  perpendicular magnetic anisotropy essential to
   all-optical ultrafast spin reversal in ferromagnets?}

%\title{Spin frustration at the core of the all-optical switching in ferrimagnets}

%\title{Paradigm of all-optical ultrafast spin switching in
%  ferromagnetic nanostructures}

\author{G. P. Zhang}

 \affiliation{Department of Physics, Indiana State University,
   Terre Haute, IN 47809, USA }

\author{Y. H. Bai}

\affiliation{Office of Information Technology, Indiana State
  University, Terre Haute, IN 47809, USA }

\author{Thomas F. George}

\affiliation{Office of the Chancellor and Center for Nanoscience
  \\Departments of Chemistry \& Biochemistry and Physics \& Astronomy
  \\University of Missouri-St. Louis, St.  Louis, MO 63121, USA }

\date{\today}

\begin{abstract}
{All-optical spin reversal presents a new opportunity for spin
  manipulations, free of a magnetic field.  Most of
  all-optical-spin-reversal ferromagnets are found to have a
  perpendicular magnetic anisotropy (PMA), but it has been unknown
  whether PMA is necessary for the spin reversal.  Here we
  theoretically investigate magnetic thin films with either PMA or
  in-plane magnetic anisotropy (IMA).  Our results show that the spin
  reversal in IMA systems is possible, but only with a longer laser
  pulse and within a narrow laser parameter region.  The spin reversal
  does not show a strong helicity dependence where the left- and
  right-circularly polarized light lead to the identical results.  By
  contrast, the spin reversal in PMA systems is robust, provided both
  the spin angular momentum and laser field are strong enough
  {\color{black} while the magnetic anisotropy itself is not too
    strong}. This explains why experimentally the majority of
  all-optical spin-reversal samples are found to have strong PMA and
  why spins in Fe nanoparticles only cant out of plane.  It is the
  laser-induced spin-orbit torque that plays a key role in the spin
  reversal.  Surprisingly, the same spin-orbit torque results in
  laser-induced spin rectification in spin-mixed configuration, a
  prediction that can be tested experimentally.  Our results clearly
  point out that PMA is essential to the spin reversal, though there
  is an opportunity for in-plane spin reversal.  }
\end{abstract}

%\pacs{75.78.Jp, 75.40.Gb, 78.20.Ls, 75.70.-i}

%ultrafast magnetization dynamics, 75.78.Jp

%75.40.Gb, 78.20.Ls, 75.70.-i, 78.47.J-}

%\keywords{femtomagnetism; exchange interaction}
 \maketitle

\section{Introduction}

Laser-controlled spin dynamics in ferromagnets started with the
pioneering work by Beaurepaire and coworkers \cite{eric}, who found
that an ultrafast laser pulse is capable of demagnetizing a nickel
thin film within 1 ps. This opens a new frontier in magnetism that has
never been seen before. Earlier studies heavily concentrated on how
such an ultrashort demagnetization occurs, a still hotly debated topic
even today {\color{black}
  \cite{koopmans2009,mathias,stamm,chan2009,turgut13,turgut16}}. Technological
implications of this discovery was recognized in the beginning, and
soon investigations of the exchange-coupled structure \cite{ju}, a
prime example in spintronics, started. The results were very
interesting, but not surprising.

A decade earlier, Gau \cite{gau} reviewed several important materials
used in magneto-optical recording (MO) and laid out the ideal
structure-property relationship for technological
applications. Amorphous rare-earth transition metals were first
reported in 1973 by the IBM group \cite{chau}, where
(Tb,Gd)$_x$(Fe,Co)$_{1-x}$ ferrimagnetic thin films with $0.2\le x\le
0.3$ represent the archetype material for MO.  These films have a
perpendicular magnetic anisotropy (PMA) and two spin
sublattices. {\color{black} The sublattices allow the compensation
  temperature, where two sublattice spins cancel each other, to be
  tuned toward room temperature \cite{gau}.}  The writing is done
  through an AlGaAs semiconductor laser and a magnetic field.

In 2007, using GdFeCo, Stanciu \et \cite{stanciu} reported that the
writing is possible even with a single laser pulse but without a
magnetic field. The helicity determines how the spin is reversed
{\color{black}\cite{elhadri2016,choi}}. Left-circularly polarized light
switches spins from down to up, while right-circularly polarized light
does the opposite. This is commonly called all-optical
helicity-dependent spin switching, or AOHDS. A group of new materials
has emerged lately \cite{lambert,mangin}. Of particular interest are
ferromagnetic CoPt ultrathin films. They only have a single spin
lattice since Pt sites have a much weaker spin moment. The films have
only 1-2 monolayers, and thicker ones do not show AOHDS.  What is
surprising is that these samples also have PMA.  { \color{black} This is
  not limited to CoPt. John and coworkers discovered that in FePt
  nanoparticles with PMA, magnetization switching is possible
  \cite{john}.  Different from CoPt and FePt, iron nanoparticles have
  an in-plane magnetic anisotropy (IMA).  Recently, Ren and
  coworkers \cite{ren} found that upon ultrafast laser excitation,
  in-plane spins in iron nanoparticles only cant out of the plane of
  their sample and are not switched over. } This indicates that the
initial spin configuration is likely to have an important impact on
how spins react to laser excitation, a theme also revealed in
dysprosium \cite{tk}.  Is PMA essential to all-optical spin switching
(AOS)? To answer this, we believe that a theoretical investigation is
imperative.

In this paper, we present a theoretical investigation to establish the
intricate connection between the spin configurations (PMA and IMA) and
switchability of the spins on the shortest possible time scale.  We
employ our newly developed model to directly compute the spin
evolution in the time domain. We demonstrate theoretically that PMA
has an unparalleled advantage over IMA to switch spins. Its switching
window is much broader, and the spin precedes orderly. The switching
is robust. By contrast, the switching in IMA systems is often subject
to chaotic spin precession. If the laser pulse is too short, the spin
reversal does not occur. Only when we increase the laser pulse
duration to 120 fs do we find a narrow region in which the spin does
reverse. The key player is the spin-orbit torque. {\color{black} At the
  optimal laser field amplitude, the in-plane torque (for IMA) is
  larger than those along the other directions, so switching
  occurs.} If we mix PMA with IMA, we find that there is no switching,
  but quite surprisingly, the spins in each layer, after the laser
  excitation, proceed in harmony, a prediction that can be verified in
  experiments. Our results clarify the role of spin configuration in
  spin reversal and should have some important implications for the
  future research in AOS.

The rest of the paper is arranged as follows. In Sec. II present our
spin reversal theory. Section III is devoted to our main results and
discussions. We start with the perpendicular magnetic anisotropy and
then move on to the in-plane magnetic anisotropy, followed by two
investigations in the mixed spin configuration.  Finally, we conclude
this paper in Sec. IV.

\section{Spin reversal theory}

Femtomagnetism \cite{ourreview} is not a traditional topic in
magnetism.  The traditional spin wave theory only describes spin
dynamics under a magnetic or thermal field \cite{prl00}, not a laser
field.  Since the beginning of AOS, enormous efforts have been made to
develop a reasonably simple theory to explain AOS, and over ten
theories have been proposed {\color{black}
\cite{stanciu,ostler,ajs,corn,khorsand,gr,mentink,berritta,kur,bar,epl16,mplb16,x.j.chen}.}

Recently, when we investigated the magneto-optical Kerr effect
\cite{epl15,mplb16}, we unexpectedly found a rather simple
method. This method only captures the initial steps of the spin
reversal, thus complementing other methods
\cite{stanciu,ostler,ajs,corn,khorsand,gr,mentink,berritta,kur,bar}
very well.  While the details of our method have been published
\cite{mplb16}, here we briefly review some basic ideas behind our
theory.  Our method starts from the traditional model, very popular in
optics and nonlinear optics, where the electron is placed in a
harmonic potential $m\Omega^2{\bf r}^2/2$ with frequency $\Omega$ and
interacts with the laser field through the dipole term, $-e{\bf
  E}(t)\cdot {\bf r}$, where $-e$ is the electron charge, ${\bf E}(t)$
is the electric field of the laser pulse, and ${\bf r}$ is the
position of the electron.  However, the traditional model has no spin,
so the interaction between the laser and spin is missing. To overcome
this deficiency, the magneto-optical theory includes a magnetic field,
${\bf B}_{ext}$, besides the electric field. The effect on the
magnetic property of the system comes from the Lorentz interaction,
but spin is still missing.  A major breakthrough came when we realized
that we can amend the spin-orbit coupling (SOC) $\lambda {\bf L}\cdot
{\bf S}$ to the original Hamiltonian, where $\lambda$ is the
spin-orbit coupling, ${\bf L}$ is the orbital angular momentum, and
${\bf S}$ is the spin angular momentum.  The electron experiences an
additional force from SOC. The orbital angular momentum is computed
from the position and momentum of the electron, i.e., ${\bf r}\times
{\bf p}$.  Then the Hamiltonian for a single electron at site $i$ is
\cite{epl15,epl16} \be H_i(t)=\frac{{\bf
    p}_i^2}{2m}+\frac{1}{2}m\Omega^2 {\bf r}_i^2 +\lambda {\bf
  L}_i\cdot {\bf S}_i -e {\bf E}(t) \cdot {\bf r}_i \label{ham1}, \ee
where the first and second terms are the kinetic and potential energy
terms, respectively, and the last term is the interaction between the
laser electric field ${\bf E}(t)$ and the system.  This Hamiltonian is
not different from those used in nonlinear optics \cite{boyd}, except
for the spin-orbit term.  This Hamiltonian is also similar to the
$t-J$ model \cite{tj}, so the itinerant nature of the electron is
captured.  But the $t-J$ model can not be used to simulate AOS, since
it has no orbital angular momentum, at least in the original
Hamiltonian where the orbital character is hidden.  What we did here
is to essentially expand the spatial dimension of the model into the
spin space. This builds a crucial link between the laser and the spin
system \cite{epl15}.  Despite this complication, being able to treat
both spin and spatial spaces opens a new path to simulate the spin
dynamics in a real time domain and permits us to attack the most
difficult issue at the hard core of AOS.

 If we fix the spin ${\bf S}$, then we recover the previous
 magneto-optical results \cite{epl15}. If we allow the spin to change,
 a source term, namely a spin-orbit torque, appears \be \left
 (\frac{d{\bf S}_i}{dt} \right )_{soc}=\lambda ({\bf L}_i \times {\bf
   S}_i)\label{spin}, \ee which describes how the spin changes at site
 $i$.  Our previous work demonstrated varieties of possible switching
 within this single-site model.  We avoid electron and spin
 temperatures on a time scale and an effective magnetic field
 \cite{ostler}.  This presents a more consistent formalism for spin
 reversal at the shortest possible time scale. For spin dynamics on a
 long time scale, one may refer to prior studies
 \cite{ostler,ajs,corn,khorsand,gr,mentink,berritta,kur,bar}.

{\color{black} Another important contribution is from the exchange
  interaction. Ramsay \et \cite{ramsay} showed that in GaMnAsP
  photocarrier spin exerts a spin-transfer torque on the magnetization
  via the exchange interaction.}  To take into account the magnetic
interaction and go beyond a single-site approximation, we include the
Heisenberg exchange interaction term for site $i$ \be H^{ex}_i=
-\sum_{j(i)}J_{ij} {\bf S}_i\cdot {\bf S}_{j},\ee where the summation
is over the nearest-neighbor site $j$ of spin {\bf S}$_i$.  $J_{ij}$
is the exchange interaction between sites $i$ and $j$, and can be
changed to simulate either ferromagnetic or antiferromagnetic
ordering. This term induces an exchange torque, \be \left(\frac{d{\bf
    S}_i}{dt}\right)_{ex}= \sum_{j(i)} J_{ij} {\bf S}_i\times {\bf
  S}_j. \label{exchange} \ee {\color{black}We also consider adding a
  magnetic anisotropy term $H_i^{\rm anis}=-d(S_{z,i})^2$, where $d$
  is the anisotropy constant. We find that for $J_{ij}=1$ eV, if $d$
  is on the order of $10^{-5}$ eV, there is no major effect on our
  results within a few hundred femtoseconds. But if $d$ is too strong,
  $10^{-3}$-$10^{-2}$ eV, the transverse components ($S_x$ and $S_y$)
  start to interfere with the longitudinal component ($S_z$), so the
  spin switching becomes difficult. This happens to the in-plane
  anisotropy case as well. In our calculation, we do not include the
  dipolar interaction since it is rather weak and only acts on a much
  longer time scale.  }  Our final Hamiltonian contains the Heisenberg
exchange term and the single site Hamiltonian \cite{jpcm11,jpcm13} \be
H=\sum_i (H_i(t)+H_i^{ex}),\label{tot} \ee where the summation is over
all the sites in the system.  To compute the spin dynamics, we
numerically solve the Heisenberg equation of motion for the spin
operator {\bf S} \cite{epl16}, $i\hbar \dot{\bf S}= [{\bf S},H]$,
where {\bf S} is an operator and $H$ is the total Hamiltonian of the
system (Eq. \ref{tot}). We employ the variable-order and variable
steps Adams method \cite{hall} to solve the differential equation. The
tolerance of calculation is set at $5\times 10^{-13}$ for over 6000
differential equations.

\section{Results and discussions}

%It is tempting to carry out a calculation on a large spin system, but
%we find that the spin dynamics is already very complicated in two-site
%systems. As a first step, studying small systems becomes very
%desirable. It will become clear below that insights gained from these
%systems are valuable to our understanding of difficult aspects of AOS.
%For this reason, in this paper we exclusively focus on a two-spin
%system.

% We assume that the system is always at 0 K since strictly
%speaking a 1D system does not have a long-range magnetic ordering.

\begin{figure}
\includegraphics[angle=0,width=1\columnwidth]{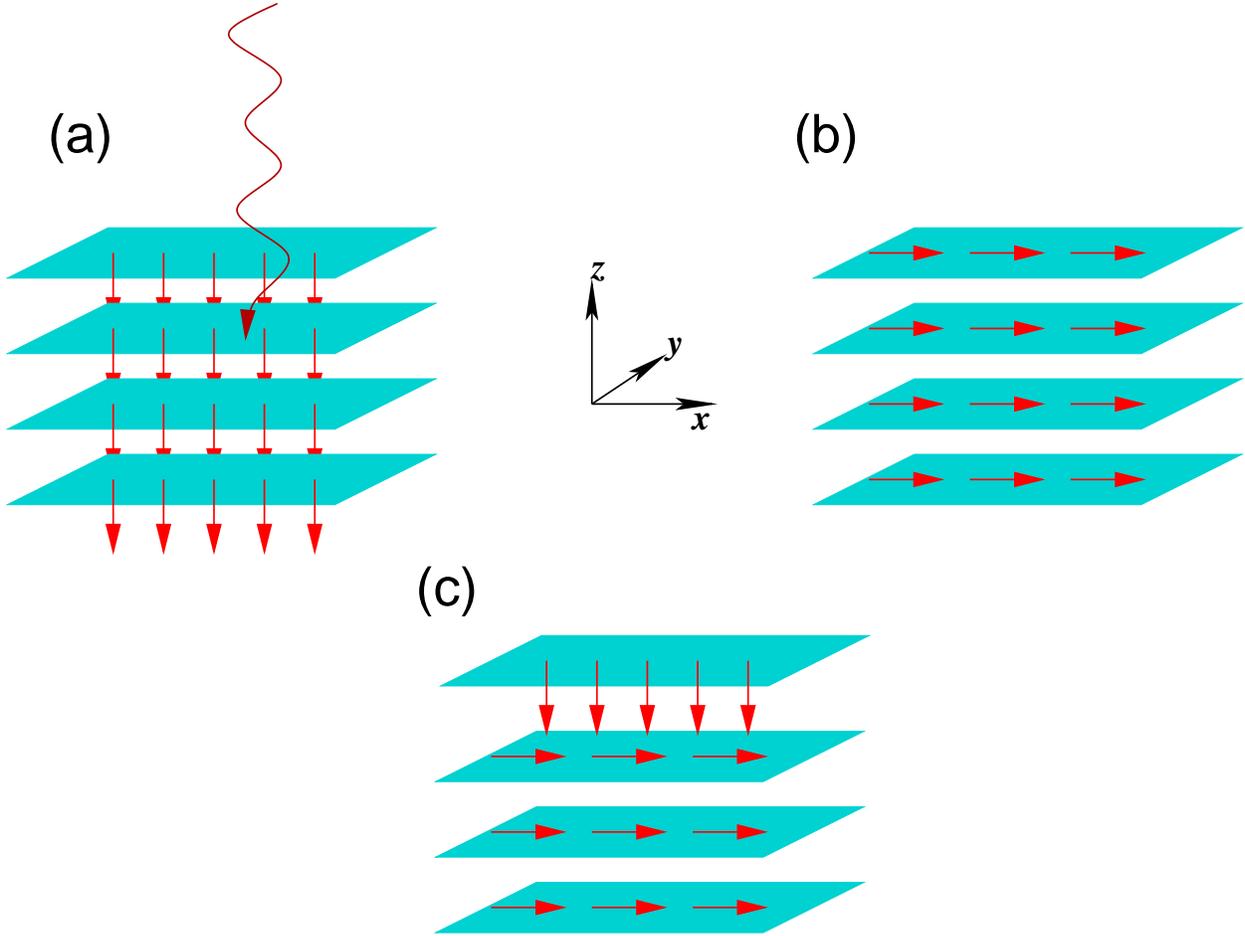}
\caption{ Three spin configurations used in this paper. Our slab has
  four monolayers and 41 lattice sites along the $x$ and $y$
  directions, with the laser light coming from the top with either
  left-circularly polarized light or right-circularly polarized
  light. The penetration depth is 40 lattice sites.  (a) Perpendicular
  anisotropy. (b) In-plane anisotropy. (c) Mixed spin
  orientation. Spins in the first layer are perpendicular, but those in
  other layers are all in-plane.  We also consider a configuration
  where the first two layers have spin out of plane and the rest in-plane.  }
\label{fig0}
\end{figure}

Our system consists of four monolayers along the $z$ direction. There
are 41 lattice sites along the $x$ and $y$ directions,
respectively. This forms a simple cubic lattice structure.  There are
over 6400 spins in our system. We have confirmed that using an even
larger number of spins has little effect on our results.  Three types
of spin configurations are considered: (a) perpendicular magnetic
anisotropy, (b) in-plane magnetic anisotropy and (c) mixed anisotropy
(see Fig. \ref{fig0}). A circularly polarized light coming down along
the $-z$ axis with a penetration depth of 40 lattice sites.
The field itself \cite{epl15,mplb16} is \be {\bf E}(t)=A_0 {\rm
  e}^{-t^2/\tau^2} (\mp \sin(\omega t) \hat{x}+\cos(\omega t)
\hat{y}),\ee where $A_0$ is the laser field amplitude, $\omega$ is the
carrier frequency, $\tau$ is the laser pulse duration, and $\hat{x}$ and
$\hat{y}$ are the unit vectors along the $x$ and $y$ directions,
respectively. $\mp$ in the equation refers to the left- (right-)circularly polarized light.  Our exchange interaction is $J=1$
eV/$\hbar^2$, and the spin-orbit coupling is $\lambda=$ 0.06
eV/$\hbar^2$, typical for transition metals \cite{prl00}.

\subsection{Perpendicular magnetic anisotropy}

We align all the spins along the $-z$ axis and couple them
ferromagnetically, with the initial value of $-1\hbar$. The laser
light impinges along the $-z$ axis. Each layer is exposed with a
different laser intensity. To start with, we use left-circularly
polarized light with duration of 60 fs.  The laser photon energy is
$\hbar\omega=\hbar\Omega$=1.6 eV. There is no magnetic field in our
simulation. Since we have 6400 spins, we decide to compute the
layer-averaged spin and also monitor each spin evolution by sampling
them individually.  Figure \ref{fig1}(a) shows the layer-averaged spin
as a function of time upon laser excitation for three components of
the spin. The solid line is the $z$ component. We see that it is
reversed successfully from -1$\hbar$ to 0.75$\hbar$. This is
consistent with our earlier calculation \cite{epl16}. {\color{black} We
  do not have full spin reversal because our current spin angular
  momentum is still too low \cite{epl16}}.  $S_x$ and $S_y$ are zero in
  the beginning, but reach around -0.5$\hbar$.  Spin oscillation is
  clearly visible for each component.  For this reason, we compute the
  maximum and minimum for each component in
  Fig. \ref{fig1}(b). Importantly, not all the laser field amplitudes
  are capable of switching the spins. Figure \ref{fig1}(b) shows the
  layer-averaged spins as a function of laser field amplitude. We
  compute the maximum and minimum spins for each component after 300
  fs (from when the laser peaks at 0 fs).  For PMA, the oscillation
  amplitude is small (compare the solid and dotted lines for $S_x$,
  $S_y$ and $S_z$).  We see that when the laser field amplitude is
  small, there is no spin reversal. But when the field increases,
  $S_z$ increases sharply, while $S_y$ decreases to a negative
  value. This change is typical \cite{mplb16} since the laser-induced
  spin-orbit torque that is needed to reverse the spin has to be
  positive along the $z$ axis, so the $S_z$ changes signs.  For a
  field amplitude that is slightly larger than 0.01 $\rm V/\AA$, the
  spin cants along the $y$ axis. A transition occurs when the laser
  field amplitude is close to 0.0165 $\rm V/\AA$ and when the spin is
  reversed along $+z$. The subtle crossing between $S_x$ and $S_y$
  signals the coming of the optimal reversal field. After this optimal
  value, the spin cants along the $-x$ axis.

One of our important findings for PMA is that it shows a strong
helicity-dependent switching. The lower line at the bottom of
Fig. \ref{fig1}(b) plots the results with right-circularly polarized
light.  We see that it can not reverse the spin in the entire
amplitude region. Here we only show the $z$ component of the spins in
the first layer, since the other components are too small to show. The huge
difference between the left- and right-circularly polarized light is
mainly due to the orbital angular momentum difference as noticed
before \cite{mplb16}.

\subsection{In-plane magnetic anisotropy}

\begin{figure}
\includegraphics[angle=0,width=1\columnwidth]{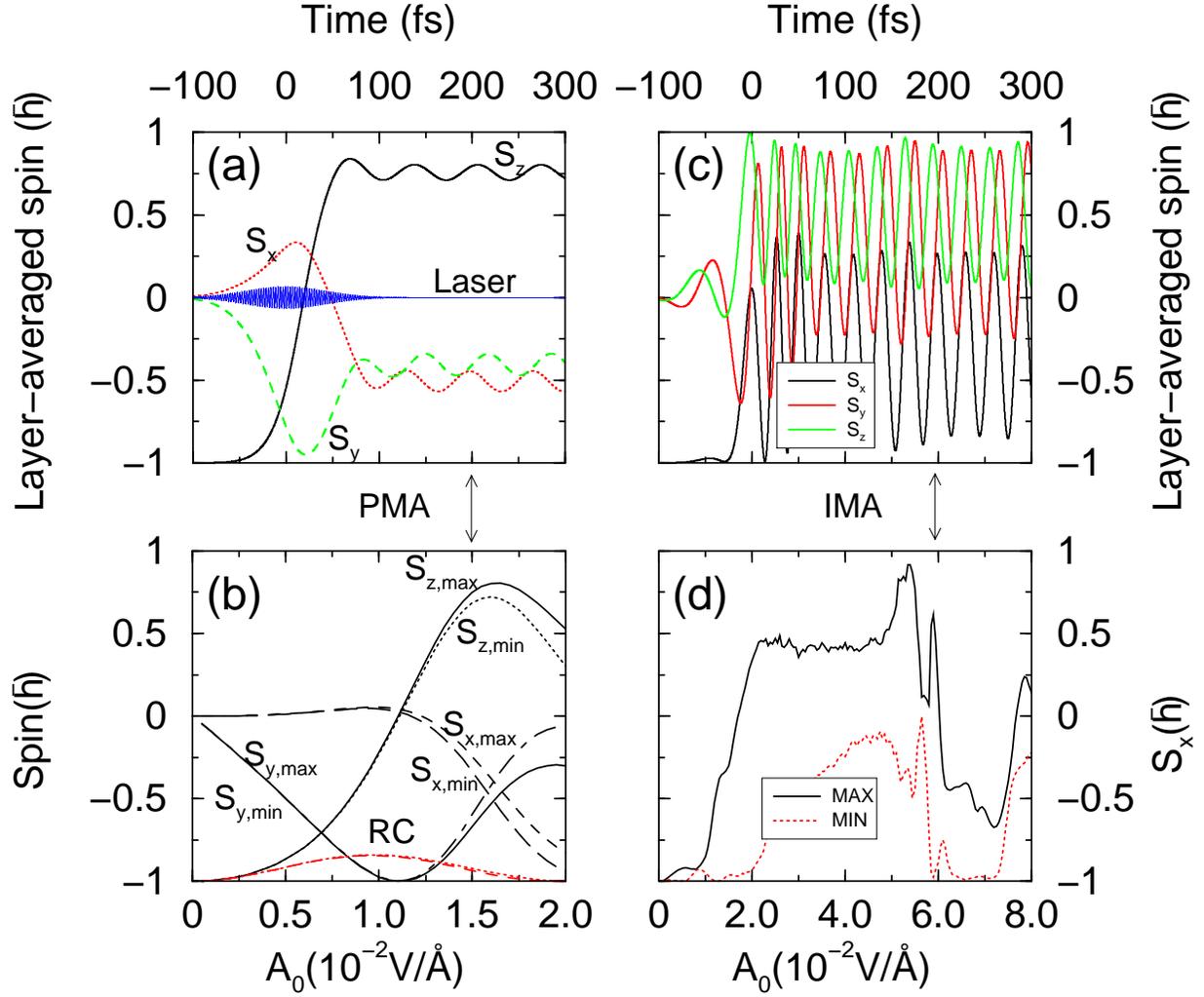}
\caption{ (a) Layer-averaged spin as a function of time for PMA. Here
  the spins are from the first layer as other layers are similar. The
  laser field amplitude is at the optimal value of 0.0165 $\rm V/\AA$
  (see (b)).  The dotted, dashed and solid lines denote the spins
  along the $x$, $y$ and $z$ axes, respectively. {\color{black} Inset: Laser
  pulse.}  (b) Laser-field amplitude
  dependence of the layer-averaged maximum and minimum spins for PMA.
  The solid and dotted lines are the spin maxima and minina along the
  $z$ axis, while rest are along either the $x$ or $y$ axis. The
  helicity-dependence is strong.  RC: the results obtained with
  right-circularly polarized light.  (c) Layer-averaged spin as a
  function of time for IMA with the spin initially along the $x$ axis.
  Here the laser field amplitude is chosen to be 0.02 $\rm V/\AA$.
  (d) Field-amplitude dependence of the layer-averaged spin along the
  $x$ axis. The maximum (solid line) and minimum (dotted line) are so
  different that a huge oscillation is found. Note that the laser
  field amplitude range is much broader than that in (b) to make sure
  that we do not miss any possible reversible window.  The laser pulse
  duration is 60 fs.}
\label{fig1}
\end{figure}

A naive guess for the in-plane magnetic anisotropy would be similar to
PMA. However, quantum mechanically, IMA has the spin quantization
along the $x$ axis. Because the optical selection rule ($\Delta l=\pm
1, \Delta m_l=0,\pm 1$ with $l$ and $m_l$, respectively, being orbital
and magnetic orbital angular momentum quantum numbers) is spatially
relative to the spin quantization, the left- and right-circularly
polarized light become equivalent to the spin. Such a convoluted
relation is difficult to include if one uses a heat pulse in place of
a true laser field. Our scheme shows the true power to simulate spin
dynamics. In Fig. \ref{fig1}(c) we present a representative result for
the spin precession. The laser field amplitude is chosen to be
0.02$\rm V/\AA$.  We find that regardless of how strong or weak the
laser field is, the switching is not observed. For a weak laser pulse,
the spin oscillates with a smaller amplitude. For a stronger laser,
the oscillation dominates the entire dynamics with a shorter
period. Therefore, the maximum and minimum spins differ a lot (see
Fig.  \ref{fig1}(d)). To be sure that we do not miss the major portion
of the laser parameter space, we extend the laser field amplitude all
the way up to 0.08 $\rm V/\AA$, and we do not find a case where the
spin is reversed with duration $\tau=60$ fs.  No spin reversal is
found, and no helicity dependence is noted.

\begin{figure}
\includegraphics[angle=0,width=1\columnwidth]{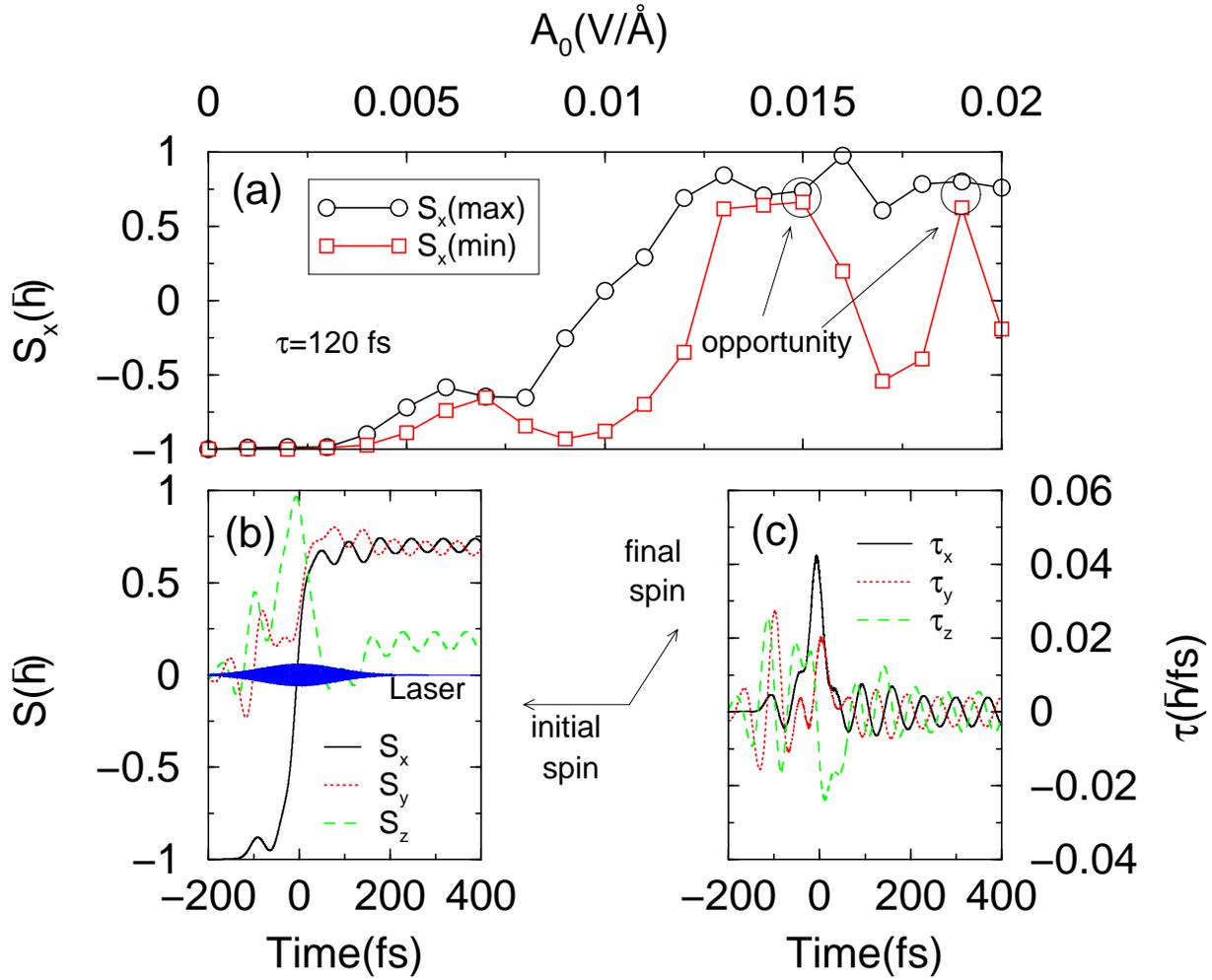}
\caption{ Longer pulses present an opportunity to reverse the in-plane
  spins.  Here a 120-fs pulse is used.  (a) Dependence of the in-plane
  component of the layer-averaged spin on the laser field
  amplitude. The empty-circle line denotes the maximum spin, while the
  empty-box line the minimum. Two circles, the first of which is
  plotted in (b), highlight two narrow regions where the reversal is
  possible. Here, the results for the first layer are shown as the
  rest are similar.  (b) Layer-averaged spin change with time. The
  solid, dotted and dashed lines denote the $x$, $y$ and $z$
  components, respectively. {\color{black} Inset in (b): Laser
    pulse.  Inset on the right of (b): Depiction of the
    initial spin and final spin orientations.}  (c) Spin-orbit torque
  as a function of time. The key insight is that $\tau_x$ is larger
  than $\tau_y$ and $\tau_z$ , although it peaks at a latter time and
  it decreases sharply once the laser pulse ends.  }
\label{fig2}
\end{figure}

From the huge oscillation in the spin in Fig. \ref{fig1}(d), we notice
something unusual.  The maximum spin can reach 0.5 $\hbar$ for the
field amplitude below 0.05$\rm V/\AA$, but its minimum is just too
negative. Considering the laser peak around 0 fs, the spin appears to
overshoot (see Fig. \ref{fig1}(c)). This is always the case with an
ultrashort pulse, where the coherence lasts very long \cite{np09}.  We
wonder whether a longer pulse can suppress such a rapid
oscillation. We increase the pulse duration to 120 fs. To our
surprise, although the entire amplitude dependence does not change
much, a window of opportunity appears. If one compares
Fig. \ref{fig2}(a) with Fig. \ref{fig1}(d), the minimum spin increases
overall, so the oscillatory amplitude drops. Because of the longer
laser pulse duration, the relative field amplitude that yields the
same change is also reduced. Around 0.015 $\rm V/\AA$, the first
optimal condition appears. Figure \ref{fig2}(b) shows the
layer-averaged spin precession as a function of time. In the
beginning, the spin is along the $-x$ axis, and {\color{black} upon
  laser excitation, it tilts within the $xy$ plane}. The $y$ component
is comparable to the $x$ component of about 0.7 $\hbar$.  In the
middle of the lower panel of Fig. \ref{fig2}, we sketch the initial
and final spins to show how the spin reversal is partially
accomplished.

To understand how the switching occurs, we need to look at the
spin-orbit torque upon laser excitation for each component of the
spin. Figure \ref{fig2}(c) shows that in general the spin-orbit torque
$\tau_{soc}$ for each component is similar, but with one crucial
exception: $\tau_{y}$ and $\tau_{z}$ peak much earlier. This
explains why $S_y$ and $S_z$ rise earlier and more quickly. This is fully
expected, since in the beginning the spins are zero along these two
directions, and the torque is a product of the spin along the other
direction with the orbital angular momentum. The challenge to
understand the spin reversal is that these three components obey the
mutual permutation relation $[S_x,S_y]=i\hbar S_z$ \cite{jpcm11}. If
one of these components is zero and stays at zero, the spin reversal
is not possible. Thus, normally the linear spin reversal proposed by
Stanciu \et \cite{stanciu} does not occur. This is reflected in our
simulation. Once $S_y$ and $S_z$ are different from zero, $S_x$ can be
switched.  Since the torques from the laser and the exchange
interaction are weak in the ferromagnetic configuration \cite{mplb16}, the
key driver to reverse $S_x$ is the spin-orbit torque. Figure
\ref{fig2}(c) further shows that $\tau_x$ (solid line) is smaller than
$\tau_y$ (dotted line) and $\tau_z$ (dashed line), except around 0 fs
when the laser peaks. Its magnitude surpasses both $\tau_y$ and
$\tau_z$. It is in this narrow temporal window that $S_x$ reverses its
direction, after which $\tau_x$ decreases. Such a decrease is necessary,
since otherwise the switched spin would undergo a strong oscillation.

In PMA, the helicity dependence is very strong, but in IMA the
helicity dependence is absent due to the selection rule discussed
above. This is also confirmed numerically.  In summary, we assert
that AOS in IMA is much harder to obtain than PMA. This explains the
crucial experimental observation as to why most AOS ferromagnets have a
perpendicular magnetic anisotropy.

\subsection{Laser-guided  spin mode rectification}

So far, our spin configurations are very pure -- either perpendicular
magnetic anisotropy or in-plane magnetic anisotropy. Magnetic
orderings can be a mix of several different configurations and may
contain different domains. Without considering all the possible spin
configurations and without changing the system size, we examine what
happens to the spin switching if the first layer of spins has PMA but
the rest IMA (see Fig. \ref{fig0}(c)). From Eq. (\ref{exchange}), we
see that such a configuration will induce a strong exchange torque, so
the spin precession is highly nonlinear. One would expect that the
laser has no big effect on the spin precession because the precession
is already chaotic. However, to our surprise, this does not happen.

\begin{figure}
\includegraphics[angle=0,width=1\columnwidth]{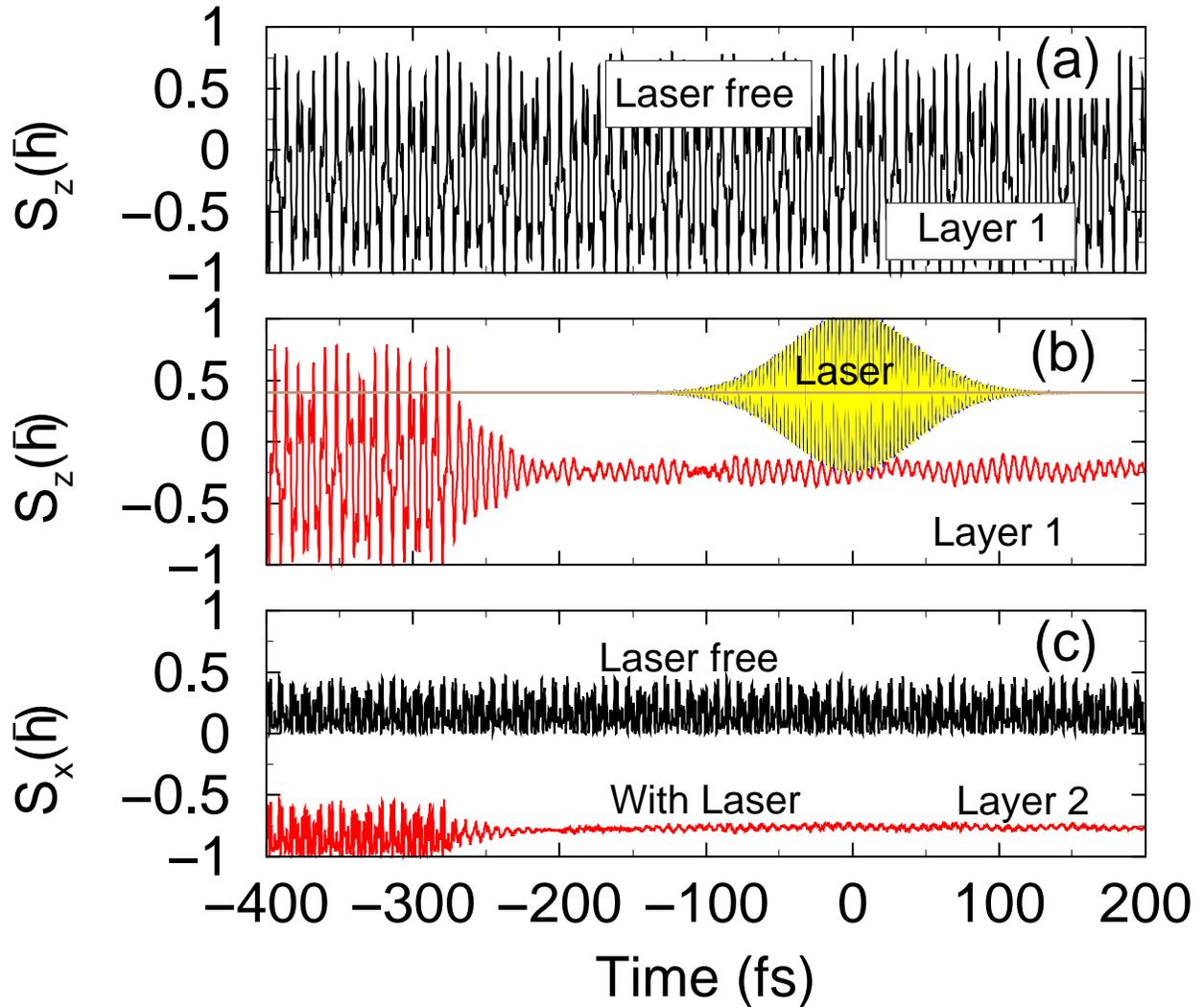}
\caption{ Laser-induced spin rectification effect in the spin-mixed
  case. The spin configuration is as follows. The first layer of spin
  points out of the plane (see \ref{fig0}(c)), while the rest are
  in-plane.  (a) A strong fluctuation of the $z$ component of the spin
  in the first layer in the absence of a laser field. (b) The same
  component as (a) but upon a 60-fs laser excitation. Inset: the laser
  pulse. The nonzero field already starts around -300 fs. (c)
  Comparison of the $x$ component of the spin in layer 2. The spin is
  also layer-averaged. On the top are the results without a laser
  field, and on the bottom those with the laser field. The top curve is
  shifted for clarity.  Other layers are similar and not shown. }
\label{fig3}
\end{figure}

We use a 60-fs laser pulse with left-circularly polarized light and
employ an extremely weak laser of $A_0=0.001\rm V/\AA$, so as to
perturb the system gently. We start with a field-free case. Figure
\ref{fig3}(a) shows $S_z$ on layer 1 changing with time. Other
components oscillate similarly and strongly overlap with $S_z$. Also
in other layers, the spins are similar, so we do not show them. This
is fully expected as discussed above.  The period of the oscillation
is determined by both the exchange interaction $J_{ij}$ and the spin
angular momentum, together with the spin-orbit coupling
\cite{mitsuko}.  Figure \ref{fig3}(b) shows the same spin component as
Fig. \ref{fig3}(a) but in the presence of a laser pulse. We see that
the fluctuation in the original field-free dynamics is strongly
suppressed. The amplitude of the spin oscillation is reduced from
$2\hbar$ to less than 0.25$\hbar$. The spin is rectified according to
the laser field.  The laser pulse is shown in the inset of
Fig. \ref{fig3}(b). Since left-circularly polarized light has nonzero
$x$ and $y$ components, these components superimpose on top of each
other in the inset. The horizontal line represents zero for the laser
field. The field has duration of 60 fs, so its nonzero field extends
from -300 fs to +300 fs.

Such spin rectification is not limited to the first layer. We find
that all the layers have this feature. Figure \ref{fig3}(c) shows the
$x$ component of the layer-averaged spin in layer 2. On the top is the
spin without laser excitation, and the bottom is with laser
excitation. The difference is very clear. We also use different laser
parameters, and the results are the same. This unexpected result
requires an experimental confirmation. Microscopically, we find that
the laser acts like a pivot that steers the spin along one
particular direction. This is consistent with our earlier finding that
the laser-induced spin-orbit torque is significantly larger than the
energy barrier to alter the spin configuration \cite{epl16}.

 To have some qualitative feeling as to how the spins change across different layers,
in Table \ref{table2} we show the initial and final layer-averaged
spins. A pattern emerges. Although the initial spins are not in
parallel, upon the laser excitation the spins in different layers
congregate into the same orientation to reduce the exchange energy.
Since there are three layers ferromagnetically coupled, the first
layer with the original PMA bends its spins toward the other layers.
Future experiments can test our prediction. Layered materials are
popular in spintronics and spin-valve devices, where one
essentially has a trilayer structure with a nonmagnetic spacer in
between two layers. If the exchange coupling is strong between two
ferromagnetic layers, even if the spin orientations at these two end
layers are different, upon laser excitation, they can be guided into the
same direction.  We also test another case where two top layers have
spin perpendicular anisotropy. As seen from the same table, the
results are similar. This suggests that this laser-induced
rectification is quite robust. We plan to investigate the helical
configuration in the future.  Experimentally, Ju \et \cite{ju} in
their first experiment did see the impact of the laser pulse on the
exchange bias. An extension of their experiment should be able to
verify our prediction.

After this work was finished, we noticed that Laliu \et \cite{lalieu}
experimentally investigated the same mechanism in noncollinear
magnetic bilayers, where one layer has spin out of plane while the
other layer spin is in plane, separated by a spacer layer. They showed
that they could absorb or generate spin currents. {\color{black} This is
  consistent with an earlier study by Huisman \et \cite{huisman2016},
  where optical generation of spin currents was demonstrated in a
  10-nm thick Co film deposited on a 2-nm Pt layer.}

\section{Conclusion}

We have investigated a hitherto open question as to whether the
perpendicular magnetic anisotropy is essential to all-optical
laser-induced spin reversal in ferromagnets. Our finding is
affirmative that PMA has a commanding advantage over IMA. AOS emerges,
as both the laser field amplitude and the spin angular momentum are
large enough. The spin does not show a strong oscillation, which is a
big advantage for future applications. However, our finding does not
exclude AOS in IMA. For a long laser pulse (120 fs), we find that
there is a narrow region where a partial reversal is still possible.
We predict that in a mixed spin configuration, a laser pulse can
effectively rectify the spin evolution by suppressing spin
frustration. We look forward to an experimental confirmation.

\acknowledgments This work was solely supported by the U.S. Department
of Energy under Contract No. DE-FG02-06ER46304. Part of the work was
done on Indiana State University's quantum cluster and
high-performance computers (obsidian and spin).  The research used
resources of the National Energy Research Scientific Computing Center,
which is supported by the Office of Science of the U.S. Department of
Energy under Contract No. DE-AC02-05CH11231.

%\vspace{1cm}

%{\bf Method}

%{\bf Numerical Simulation }

%\vspace{1cm}

%\clearpage

\clearpage

\begin{table}
%original data on spin: /home/gpzhang/doe/switch/config/job/41x41x4/mixed/LC/spin-100.60fs
%original data on obsidian: /net/home/gzhang/doe/switch/config/job/41x41x4/mix2layer/spin-100.60fs
\caption{ Spin change from the initial $S^i$ to the final $S^f$ for
  two
mixed spin configurations under laser excitation. The laser pulse
duration is 60 fs. $A_0=0.001\rm V/\AA$. All the final spins are
collected at time $t=661.86$ fs.  The numbers in  parenthesis are
the $x$, $y $ and $z$ components. The two columns represent two different
spins configurations (see Fig. \ref{fig0}). On the left, the spins in the
first layer point along the $-z$  axis, and the rest are in-plane; On the
right, the spins in the first two layers point along the $-z$ axis, and the rest are in-plane.
 }
\vspace{1cm}
\begin{tabular}{ccc|cc}
\hline\hline
Spin$\Longrightarrow$ &
\multicolumn{2}{c}{$\downarrow\leftarrow\leftarrow\leftarrow$}
&\multicolumn{2}{c}{ $\downarrow\downarrow\leftarrow\leftarrow$}
\\
%Layer & ~~~Initial spin ($\hbar$) ~~~      & ~~~ Final spin ($\hbar$)
%~~~  &~~~Initial spin ($\hbar$) ~~~      & ~~~ Final spin ($\hbar$)
%~~~  \\
\hline
Layer &              $S^i(\hbar)$       & $S^f(\hbar)$
     &$S^i(\hbar)$      & $S^f(\hbar)$
  \\
\hline
%filename:results.Nspin.for.layer01.Intensity0.00100
1 & (0,0,-1)       &  (-0.76,    0.08, -0.20)& (0,0,-1)       &  (-0.47, -0.04, -0.53)\\
%filename:results.Nspin.for.layer02.Intensity0.00100
2 & (-1,0,0)       &  (-0.77,    0.03, -0.22)& (0,0,-1)       & (-0.49, -0.04, -0.52)\\
%filename:results.Nspin.for.layer03.Intensity0.00100
3 & (-1,0,0)       &  (-0.77,   -0.05, -0.23)& (-1,0,0)       & (-0.52, -0.03, -0.49)\\
%filename:results.Nspin.for.layer04.Intensity0.00100
4 & (-1,0,0)       &  (-0.75,   -0.01, -0.23)& (-1,0,0)       & (-0.53,  -0.04, -0.46)\\
\hline\hline
\label{table2}
\end{tabular}
\end{table}

%\clearpage

%\onecolumngrid

%In spin-transfer torque (STT), the switching speed depends on the
%angular momentum transfer from the current to magnetic layers
%\cite{koch,bedau}.
%\clearpage

\end{document}